\documentclass{webofc}

\usepackage[varg]{txfonts}   
\usepackage{hyperref}
\usepackage{url}

\hypersetup{colorlinks=true,citecolor=blue,urlcolor=blue,linkcolor=blue}
\begin{document}
\title{Cumulants of the multiplicity distributions of identified particles measured in heavy-ion collisions by HADES}
%
%

\author{\firstname{Marvin} \lastname{Nabroth for the HADES Collaboration}\inst{1}\fnsep\thanks{\email{m.nabroth@gsi.de}} 
}

\institute{Institut für Kernphysik, Frankfurt}

\abstract{The HADES experiment investigates the reaction products of heavy-ion collisions at a few GeV, providing access to QCD matter at high net-baryon densities. A particular focus is the reconstruction of higher-order cumulant ratios of proton and light nuclei multiplicity distributions, which are considered sensitive probes of criticality through their connection to event-by-event (E-by-E) net-baryon number fluctuations. In this contribution, we summarize the experimental setup and outline the analysis strategy. The fluctuations are reconstructed using a novel approach that circumvents E-by-E particle identification by treating it probabilistically. We present several approaches for efficiency correction and apply a data-driven event-mixing procedure to correct for centrality fluctuations. The fully corrected normalized factorial cumulants of proton and deuteron multiplicity distributions are then presented and compared to corresponding STAR data. Our measurements extend the trends observed by STAR toward lower energies for the ratios $C_2/C_1$ and $C_3/C_1$. Finally, we show the rapidity dependence of factorial cumulant ratios and confront the results with canonical ensemble baselines that incorporate correlated particle production via local attractive interactions.
}

\maketitle
\section{Physics Motivation}
\label{intro}
Low-energy heavy-ion collisions in the few GeV region generate, for a short period of time, QCD matter at high net-baryon densities \cite{HADES:2019auv}. This provides access to the region of the QCD phase diagram where a first-order phase transition and a critical end point are expected. E-by-E fluctuations of the net-baryon number should be reflected in the emitted numbers of protons and light nuclei and are quantified by extracting their cumulants $\kappa_{n}$ of order $n$, which, in the context of statistical physics, are linked to derivatives of the Grand Canonical partition function w.r.t. baryon-chemical potential. To reveal potential signatures of criticality, the objective is to study regular or factorial cumulant ratios \footnote{Linear combination of $\kappa_{n}$} as a function of collision energy. A non-monotonic dependence of these observables has been proposed as a possible indicator of critical phenomena \cite{Stephanov:2011zz, Cheng:2008zh}. Before drawing any physical conclusions, it is essential to investigate and correct for non-critical sources of fluctuations caused by detector effects and reaction-volume variations. Furthermore, the observed trends must be interpreted relative to baseline expectations, to distinguish genuine critical signals from background effects.

\section{Experimental Setup and Analysis procedure}
HADES operates at the SIS18 synchrotron at GSI Darmstadt \cite{HADES:2009aat}. The detector is designed as a fixed-target experiment, allowing for nearly complete azimuthal coverage. The analysis presented in this contribution is based on the measurement campaign of Ag+Ag collisions at $\sqrt{s_{NN}}=$ $2.55$ $GeV$, conducted in March 2019. Ag beam ions are detected by the START detector before interacting with a 15-fold segmented target. The reaction products from the collisions are boosted into the spectrometer, which is divided into six identical sectors. Before entering the magnetic field region, the particle tracks are measured by two inner Multi-Wire Drift Chambers (MDC). In the magnetic field region, the particle trajectories are deflected, and their positions are subsequently measured behind the field region by two outer MDCs. Behind the outer MDCs, two Time-of-Flight detectors, RPC and TOF, part of the META system, are located. These detectors have the purpose of determining the time of flight, in combination with the reference time registered by the START detector. The momentum-to-charge ratio is derived from the measured deflection in the magnetic field region. Reconstructed track candidates are then matched with hits in the META detectors \cite{HADES:2009aat}. The time-of-flight measurement, together with the reconstructed momentum, provides a direct access to the particle's mass to charge ratio, which serves as the primary observable for particle identification (PID) in this analysis. Additionally, the MDCs provide measurements of the specific energy loss (dE/dx), which are used in the analysis as a pre-selection criterion.  

\begin{figure}[h]
		\centering
		\begin{minipage}[c]{0.5\textwidth}
			\includegraphics[width=\textwidth]{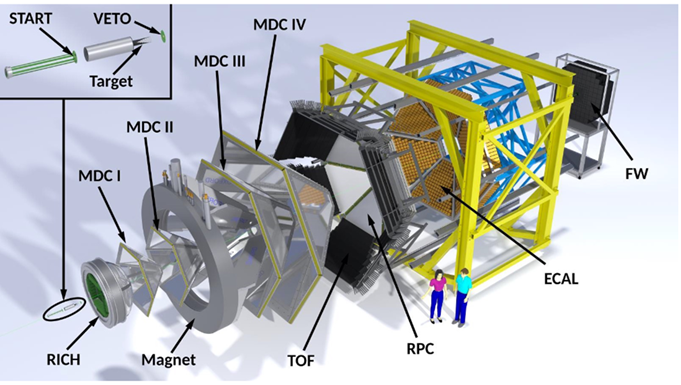}
		\end{minipage}
		\begin{minipage}[c]{0.43\textwidth}
			\includegraphics[width=\textwidth]{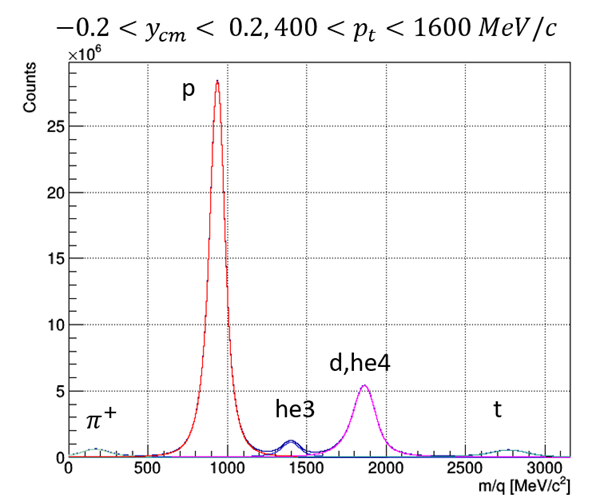}
		\end{minipage}
		\caption{Left: Explosion view of the HADES detector with its main detector components. Right: Integrated mass spectra line shapes for the selection region: $400 < p_t < 1600$ $MeV/c$, $-0.4 < y_{cm} < 0.4$.}
		\label{fig_deteector_PID}
\end{figure} 

\subsection{Event and centrality selection}
For the E-by-E dependent analysis, a clean and high-quality sample of Ag+Ag events is crucial. Various selection criteria are applied to filter out events originating from outside the target region, events with degraded time determination, or those contaminated by reactions in the detector material. The selection of collision centrality is based on the measurement of the projectile spectators using a Forward Wall hodoscope placed behind the main spectrometer \cite{HADES:2009aat}. The use of this centrality estimator ensures a weak correlation with protons, which is essential, as a selection based on charged tracks or detector hits would significantly bias the proton distribution.
\subsection{Particle identification}
The proton sample is selected in a defined phase-space region of $400$ $MeV/c$ $<p_{t}<$ $1600$ $MeV/c$, $-0.4 < y_{cm} < 0.4$. Where $p_t$ corresponds to transverse momentum and $y_{cm}$ to rapidity in the center of mass system. For each phase space bin, as well as for each of the six detector sectors, the mass spectra are fitted using six different modified two-sided Crystal Ball functions \cite{RefJ}. The aim is to describe the full inclusive spectrum with a combined fit. The resulting fit functions can be used either for a hard-cut selection or for a probability-based particle identification (PID) procedure, leveraging the principles of fuzzy logic \cite{Rustamov:2024hvq}. The basic idea is to calculate for each particle candidate the probability from the fit function. The probability sum $W$ over all tracks then serves as a proxy for the particle number. It is possible to express the relationship between the moments of the probability proxy $<W^m>$ and the moments corrected for misidentification $<N^m>$ as a regular matrix, allowing to get the true moments by simple matrix inversion. \cite{Rustamov:2024hvq}.

\subsection{Efficiency corrections}
\begin{figure}[h]
		\centering
		\begin{minipage}[c]{0.45\textwidth}
			\includegraphics[width=\textwidth]{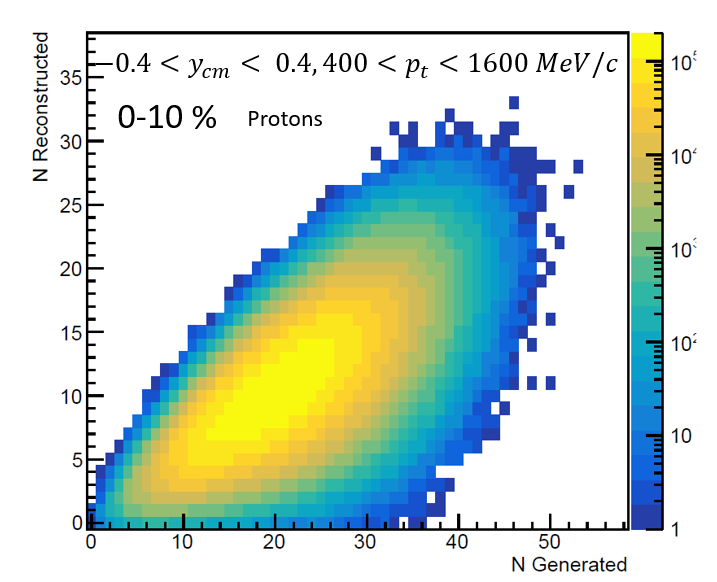}
		\end{minipage}
		\begin{minipage}[c]{0.5\textwidth}
			\includegraphics[width=\textwidth]{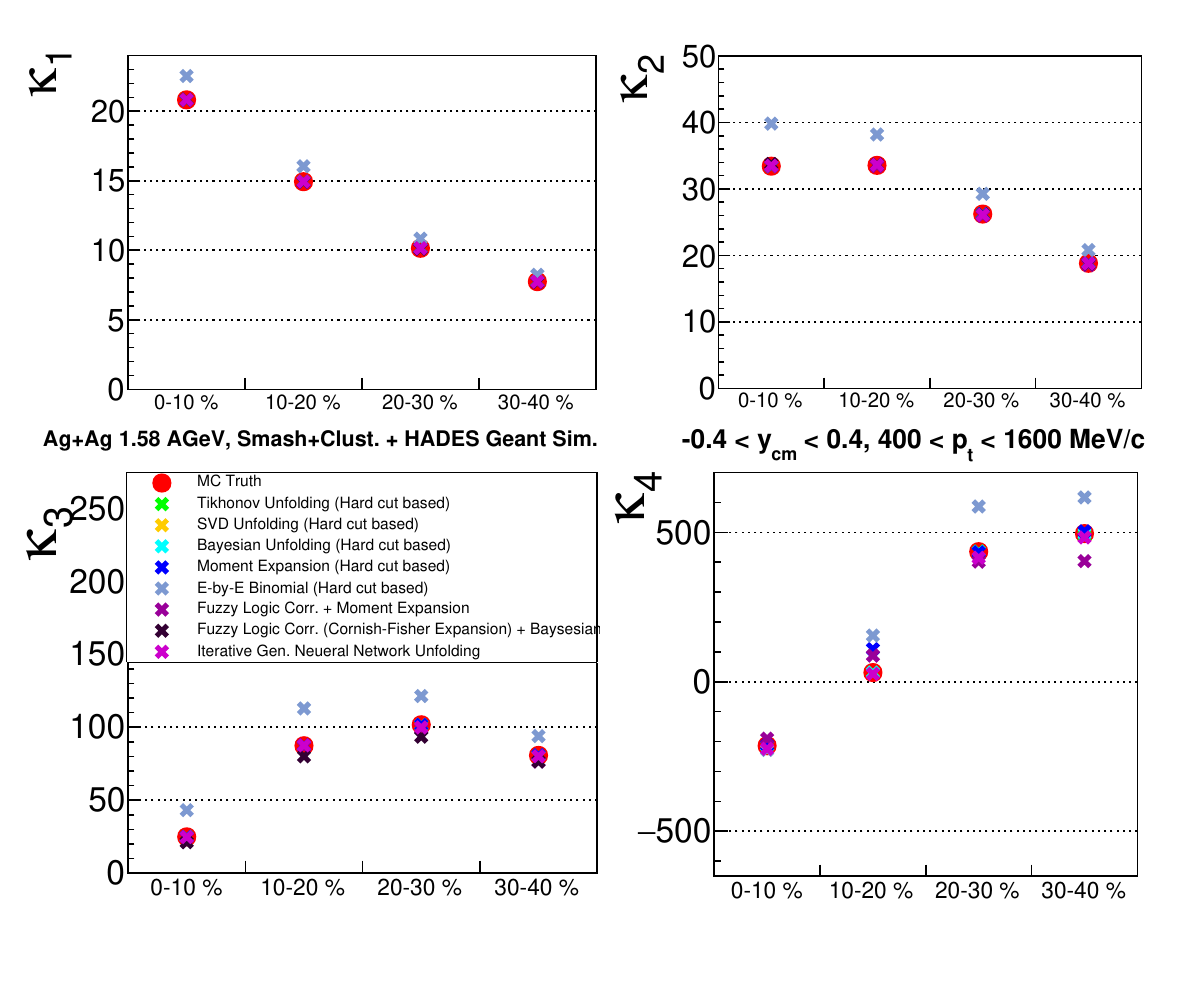}
		\end{minipage}
		
		\caption{Left: Detector repose matrix depicting the relation between the number of MC (generated) protons and the number of reconstructed/identified protons. Right: Closure-Test - Unfolded proton cumulants in comparisons with the cumulants of the MC Truth.}
		\label{fig_eff}
\end{figure}

The selected proton sample is then corrected for limited detector efficiencies. Various methods based on Binomial Correction \cite{Nonaka:2017kko}, distribution unfolding \cite{RefEff1} \cite{Schmitt:2012kp}, and direct correction of moments \cite{Nonaka:2018mgw} were evaluated in a Monte Carlo (MC) closure test. The test setup consists of a transport simulation combined with a clustering algorithm. The particles are propagated through a virtual detector setup using the GEANT package to emulate the interaction of the particles with the detector material, as well as the track reconstruction process as employed on real data.
For the event-by-event (E-by-E) binomial correction, a parameterization of the efficiency as a function of phase space, sector, and track density is required. For the other approaches, the detector response matrix is the foundation, see Fig. \ref{fig_eff}), which describes the correlation between the MC truth and the number of reconstructed particles.
It was found that all binomial correction methods fail the MC closure test. In contrast, distribution unfolding methods and direct correction of moments pass the closure test and are therefore applied to real data. The reconstruction performance in comparison with the MC truth is shown in the right panel of Fig. \ref{fig_eff}.

\subsection{Controlling volume fluctuations}
The next step in the analysis chain is to correct for fluctuations of the reaction volume. This is achieved using a fully data-driven event-mixing approach, as proposed in \cite{Holzmann:2024wyd, Rustamov:2022sqm}. The method aims to produce Poisson-like particle distributions after event-mixing, from which the volume cumulants are determined within the framework of the Wounded Nucleon Model \cite{Braun-Munzinger:2016yjz}. The assumption of Poisson-like particle emission is supported by an MC-Glauber fit to the charged-track distribution, which yields an emission function consistent with a negative binomial distribution close to a Poissonian.

\section{Results}
\begin{figure}[h]
		\centering
		\begin{minipage}[c]{0.4\textwidth}
			\includegraphics[width=\textwidth]{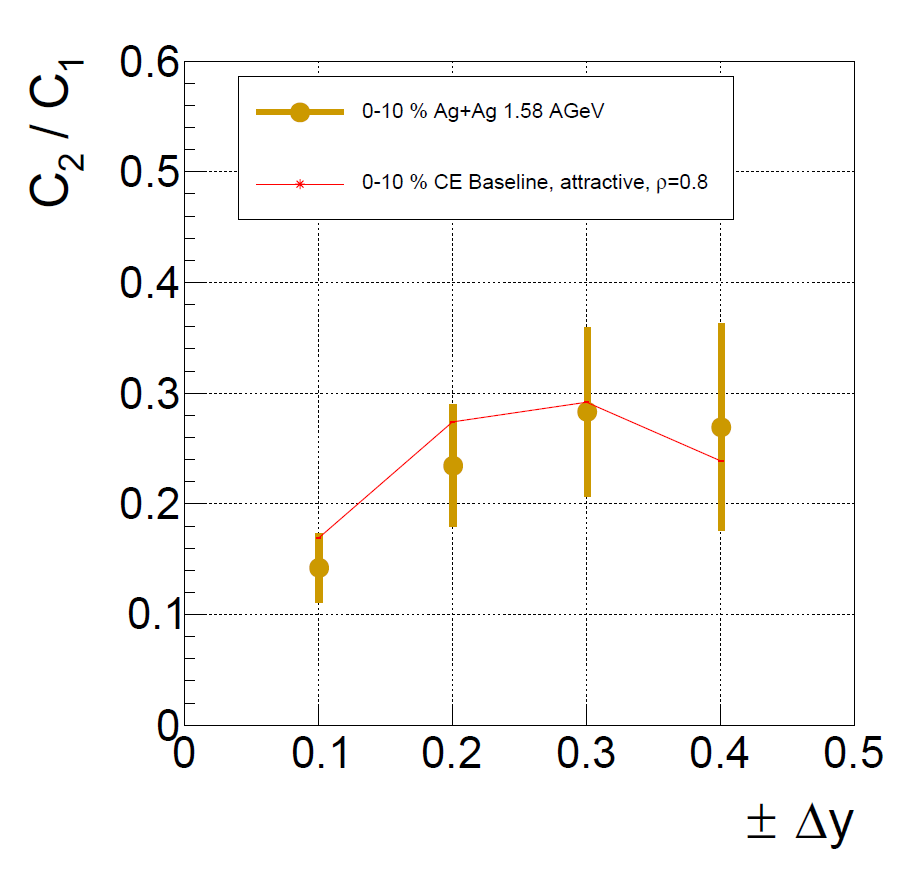}
		\end{minipage}
		\begin{minipage}[c]{0.43\textwidth}
			\includegraphics[width=\textwidth]{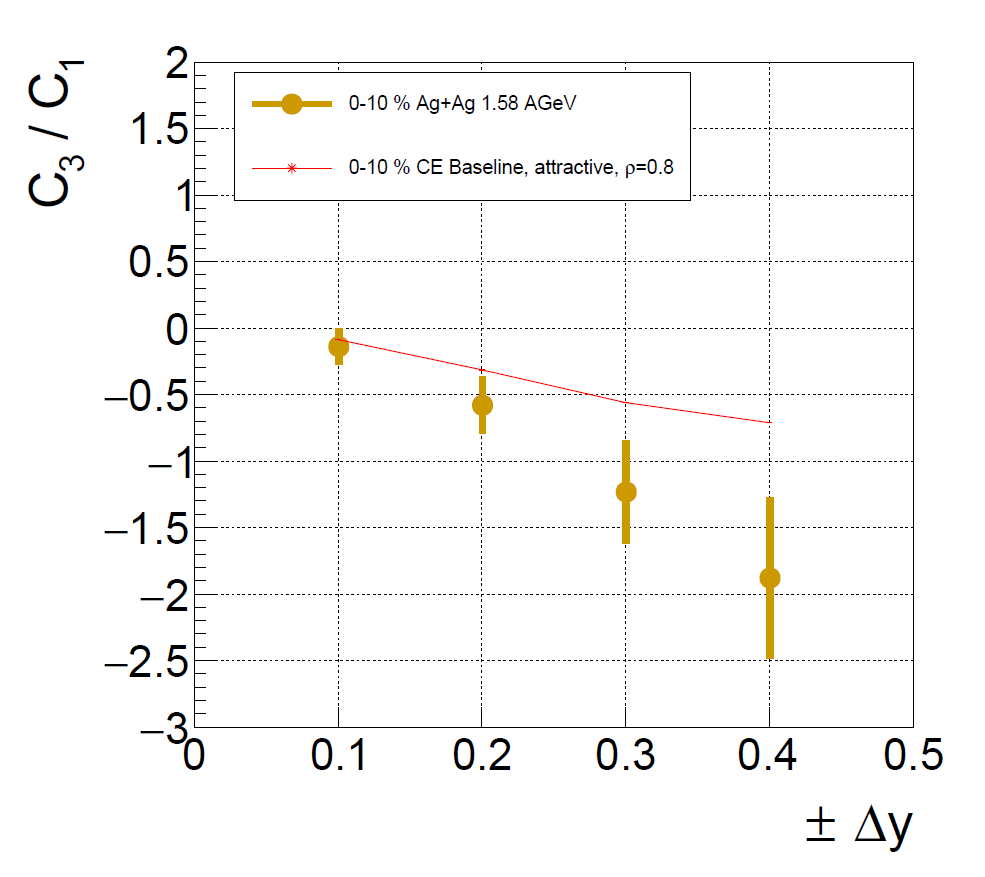}
		\end{minipage}
		\caption{Proton factorial cumulants as a function of rapidity acceptance w.r.t. mid-rapidity, for the pt-range ( $400$ $MeV/c$ $<p_{t}<$ $1600$ $MeV/c$). Red lines corresponds to canonical ensemble with correlated particle production based on an attractive potential with correlation coefficient $\rho$ \cite{Friman:2025swg}. }
		\label{fig_faccum}
\end{figure}

Figure \ref{fig_faccum} shows the efficiency- and volume-corrected proton factorial cumulants as a function of the rapidity acceptance window around mid-rapidity. In general, with decreasing rapidity acceptance, we observe a convergence towards the small number Poisson limit. The functional dependence cannot be described by a simple canonical baseline \cite{Braun-Munzinger:2020jbk}. Instead, it is found that a more elaborate canonical ensemble model that takes into account correlated particle production with an attractive potential describes the trend across all orders \cite{Friman:2025swg}. The correlations were included using the Metropolis algorithm with initial random numbers drawn from fits to the reconstructed rapidity distributions \cite{Friman:2025swg}. In comparison with STAR data, the HADES points continue the trend observed toward lower energies. However, recently presented results from STAR at 3.3 GeV indicate a negative sign for $C_4/C_1$ \cite{QMProceedingSTAR}. The opposite trend observed by HADES may be attributed to differences in acceptance windows, collision systems, and the treatment of volume-fluctuation corrections (mixed events versus Centrality Bin Width Correction \cite{Luo:2013bmi}).

\section{Outlook}
The fuzzy-logic-based correction scheme for particle misidentification has also enabled the analysis of deuterons. A proper baseline will be essential in future to fully interpret the observed trends in deuteron cumulants. In addition, the fuzzy-logic approach facilitates the reconstruction of mixed moments, offering new opportunities for particle-correlation studies.
Up to now, the volume correction has relied on the Wounded Nucleon Model, which assumes that protons are emitted from statistically independent sources. A next step is to combine the event-mixing method with a modified volume model in which this assumption is not a prerequisite, e.g. following the ansatz proposed in \cite{HADES:2020wpc}.

\bibliography{ref}

\end{document}